# THE ENERGY NAVIGATOR
## A WEB BASED PLATFORM FOR FUNCTIONAL QUALITY MANAGEMENT IN BUILDINGS

Stefan Plesser[1,3], Norbert M. Fisch[1],
Claas Pinkernell[2,3], Thomas Kurpick[2], Bernhard Rumpe[2],

1 Institute of Building Services and Energy Design, Braunschweig University of Technology, Germany
{fisch, plesser}@igs.bau.tu-bs.de
http://www.igs.bau.tu-bs.de

2 Software Engineering, RWTH Aachen University, Germany
{pinkernell, kurpick, rumpe}@se-rwth.de
http://www.se-rwth.de

3 synavision GmbH, Aachen, Germany
{pinkernell, plesser}@synavision.de
http://www.synavision.de

ABSTRACT
Energy efficient buildings require high quality standards for all their technical equipment to enable their efficient and successful operation and management. Building simulations enable engineers to design integrated HVAC systems with complex building automation systems to control all their technical functions. Numerous studies show that especially these supposedly innovative buildings often do not reach their energy efficiency targets when in operation. Key reasons for the suboptimal performance are imprecise functional descriptions and a lack of commissioning and monitoring of the technical systems that leave suboptimal operation undetected. In the research project "Energy Navigator" we create a web-based platform that enables engineers to create a comprehensive and precise functional description for the buildings services. The system reuses this functional description – written in an appropriate domain specific language – to control the building operation, to signal malfunctions or faults, and in particular to measure energy efficiency over time. The innovative approach of the platform is the combination of design and control within one artifact linking the phases of design and operation and improving the cost effectiveness for both services. The paper will describe the concept of the platform, the technical innovation and first application examples of the research project.

## 1. INTRODUCTION

Energy efficiency of buildings depends on their hardware – their construction, insulation, glazing, HVAC systems etc. – and on the way they are used and operated. While politics and technical guidelines as well as the historically developed procedures focused on hardware to improve energy efficiency, the software – user awareness, monitoring, commissioning – is now being moved into the spot light in Germany.

Technical guidelines like DIN EN 15232[1] help to evaluate the potential increase in energy efficiency by using new building management technologies and algorithms. The guideline defines different BMS solutions that allow an easy A-D classification for different building services with a corresponding saving potential percentage.

Beyond the building design for the first time mandatory codes such as EnEV 2009[2] require not only a certain level of energy efficiency on the basis of a calculated energy demand using models like defined by DIN V 18599[3]: During the life cycle large air conditioning systems have to be inspected regularly every 10 years.

---

[1] DIN 15232: Energieeffizienz von Gebäuden - Einfluss von Gebäudeautomation und Gebäudemanagement; Deutsche Fassung EN 15232:2007-11
[2] EnEV 2009: Energieeinsparverordnung 2009
[3] DIN V 18599: Energetische Bewertung von Gebäuden - Berechnung des Nutz-, End- und Primärenergiebedarfs für Heizung, Kühlung, Lüftung, Trinkwarmwasser und Beleuchtung, 2007-02

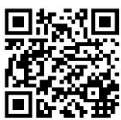

DIN EN 16001[4] goes beyond these technical recommendations by defining an overall controlling process for building owners, public administration or businesses. It helps to establish a process of installing a monitoring system, gathering and analysis of data, reporting, defining the responsibilities within the institution and the documentation of savings. The whole process is supposed to be part of a wider sustainability strategy. Its implementation is a criterion to receive tax reduction thereby providing cost savings even without immediate cuts in energy consumption.

Although commissioning is strongly regulated by German building laws and defined by technical guidelines like VDI 6022[5], DIN 15239[6], 15240[7] and others, it is further strengthened by certification labels by the German *DGNB – Deutsche Gesellschaft für Nachhaltiges Bauen* or the *US American LEED-Standard*. Both require commissioning and monitoring actions to improve building operations including functional performance testing.

The target of any commissioning and monitoring regarding building operations is to optimize the functions of the building in operation. Although it is obvious that functions in operation must have their roots in the functional design of buildings the majority of publications in this field focuses on BMS data analysis, fault detection in operation and successful re-commissioning in the later operations. A good survey on approaches is given by Katipamula[8] and the IEA Annex 47 reports[9]. There is no research work to the knowledge of the authors that focuses explicitly on the way a functional description is actually being used after the design phase to monitor operations and serve as adaptable part of the building documentation. This aspect becomes even more important when innovative buildings undergo comprehensive operational adjustments in the first year of operation.

Crucial for the large scale success of monitoring buildings in operation as well as for stronger political measures requiring monitoring are cost effective solutions. Therefore the starting point for the Energy Navigator was not the technical opportunities that the internet and BMS provide to analyze data. The innovative approach started not by looking for what is possible but for what is needed to ensure good building performance.

2. QUALITY MANAGEMENT APPROACH

The authors have carried out numerous analyses of buildings in operation[10]. One of the conclusions of these projects and point of origin for the Energy Navigator is that energy efficient operation requires cost effective solutions for

 a. Quality control by monitoring using a model of the buildings functions and
 b. Design experts to set up and carry out the monitoring services.

The authors' hypothesis is that monitoring can only be successfully integrated if it starts in the design process in which the design engineer defines the major functions of the building systems.

According to the German state of the art engineering process the design engineer describes the functions of heating, cooling and ventilation system according to VDI 3814[11] (respectively DIN EN 16848[12]), Figure 1. These guidelines describe the general functions of building automation systems and provide rules for the description of their individual application in buildings. They focus on hardware datapoints, actors and sensors and basic functions of building automation systems like limit checking, event counting or command execution check. Figure 2 and 3 show an example of a hot water circuit with a control valve and the structure of the Active Functional Description.

---

[4] DIN EN 16001: „Energiemanagementsysteme", 2009-08

[5] VDI 6022: Technische Regel, Hygiene-Anforderungen an Raumlufttechnische Anlagen und Geräte, 2006-04

[6] DIN EN 15239: Lüftung von Gebäuden - Gesamtenergieeffizienz von Gebäuden - Leitlinien für die Inspektion von Lüftungsanlagen; Deutsche Fassung EN 15239:2007

[7] DIN EN 15240: Lüftung von Gebäuden - Gesamtenergieeffizienz von Gebäuden - Leitlinien für die Inspektion von Klimaanlagen; Deutsche Fassung EN 15240:2007

[8] Katipamula, S, Brambley, M.: Methods for Fault Detection, Diagnostics, and Prognostics for Building Systems-A Review, Part I, HVAC&R RESEARCH, VOLUME 11, NUMBER 1, 2005-01

[9] See http://www.iea-annex47.org/

[10] Plesser et alt: EVA – Evaluation of Energy Concepts for Office Buildings", Technische Universität Braunschweig, 2008

[11] VDI 3814: Gebäudeautomation, Verein Deutscher Ingenieure, May 2005

[12] DIN EN ISO 16484-1: Systeme der Gebäudeautomation, May 2009-05

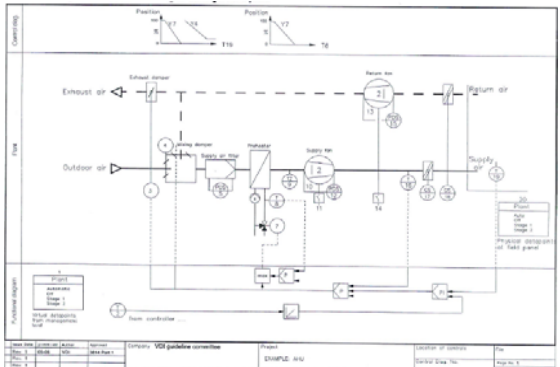

**Figure 1 Automation Scheme as defined by VDI 3814-1 (example for central air handling unit)**

The guidelines lack an explicit concept to define control loops, set points and time schedules by design engineers. Instead existing design tools focus largely on the hardware description to sum up data points, size switchboards etc. for cost calculation. In practice the actual design of the systems function is left to free textual description with the obvious consequences of incompleteness and inconsistencies.

As a solution the Energy Navigator creates a Domain Specific Language that allows a standardized and explicit description of building functions that serves three functions:
   a. Clear-cut definition of building functions as part of the building design and to be included in tenders for manufacturers or installation contractors.
   b. Model for building functions as the basis for monitoring services.
   c. Adaptive documentation after initial commissioning of the building and optimization.

The same functional definitions serve as a design document in the tender as well as for the monitoring of the corresponding functions using historical data of the building management system. The functional design becomes the model of the building functions and can therefore be used as reference model to evaluate operations.

All functions can be supplemented by variables for the degree of fulfillment of the functional design intend in operation. These variables, called "Betriebsgüte" (functional performance), can also be described in the tender. As a consequence performance quality becomes a clear-cut and owed service that can be measured by the degree of functional fulfillment.

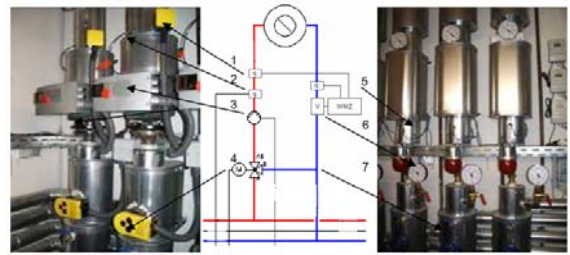

1: Supply temperature sensor heat meter
2: Supply temperature sensor,
3: Electronic pump
4: Motor valve,
5: Return temperature sensor heat meter,
6: flow meter,
7: return water admixing

**Figure 2 Hot water loop for radiant heating**

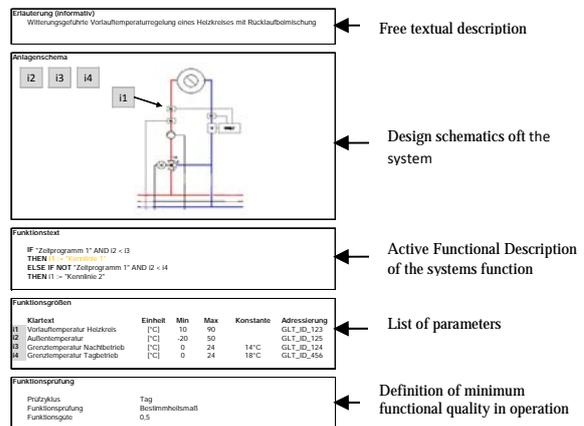

**Figure 3 Structure of an Active Functional Description for a hot water circuit**

All parts of this concept are worked out on an internet platform. Therefore the approach followed here has several advantages:
   a. The modeling of a building that is usually done in addition to the design as part of an additional and expensive monitoring service is dispensable.
   b. All partners – design engineers, installation contractors, monitoring experts, building owners - work on the same platform.
   c. Design and monitoring of functions use the same rules guaranteeing responsibilities of engineers and manufacturers.
   d. Rules can be adapted throughout the lifecycle and kept up to date.

By providing a uniform model to describe and control building functions the Energy Navigator closes the gap between design and operation.

## 3. TECHNICAL REALIZATION

The Energy-Navigator Platform is divided into a client side that can be web based or desktop based. The web based client concentrates on visualization of reports for end users of a building (management reports, public awareness). The rich client targets the energy experts that specify, evaluate and validate the proper operation of a building. The platform is designed for multiuser access.

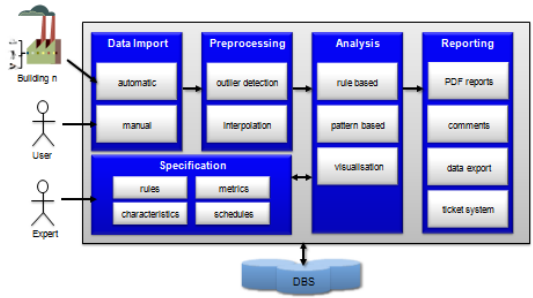

**Figure 4 Software architecture of Energy Navigator platform**

On the server side there are several components forming the backend platform. Each component can be plugged in, so that the platform is highly adoptable to build complex solution for each building. A brief overview is given in Figure 4. The components are grouped into Data Import, Preprocessing, Analysis, Report and Specification. The platform offers automatic or manual (file based) import depending on the data availability inside the building system. After the import data can be automatically preprocessed. This is a crucial step to guarantee a high data quality for analysis and further calculations. There is outliner detection or interpolation of missing values, a transformation to equidistant time steps, to mention only some preprocessing workflows. The analysis can be done automatically by using a formal specification model and intelligent algorithms. Additionally an energy expert can use the platform to visualize the imported data with multiple plot types, e.g. line plots, scatter plots or carpet plots. Another feature of the Energy Navigator Platform is the reporting component. With this component we are able to automatically create reports and inform other users about the current system status, and potential performance issues of the building.

Another feature of our platform is the possibility to formalize the knowledge of an energy expert for specifications of buildings, facilities or cause-effect relationships. An energy expert can use rules, metrics, characteristics or time routines to formalize his/her knowledge that can be automatically evaluated by our system.

The platform uses a database backend and high efficient algorithms that are optimized to handle mass data of current building systems. The data can be processed and stored localized, e.g. for reasons of data privacy protection, or with innovative cloud computing technologies for better scalability and resource sharing.

A crucial aspect is how to deal with complexity in the design of buildings. The Energy Navigator platform establishes the concept of templates for every artifact. The idea behind this concept is that an expert can specify his/her knowledge once at the beginning and use these templates easily for every building that he/she operates or manages. For a concrete building the expert adds the templates to the workspace and maps the concrete sensors to the template. The use of a library is a key feature for reusability of expert knowledge.

### Precise Constraint Language
The main concept of the Energy Navigator platform is a constraint language that is used to specify facilities, systems and building operation. The constraint language is an adapted variant of the Object Constraint Language (OCL), part of the well known Unified Modeling Language (UML).[13] The language is developed with MontiCore[14] - a framework for efficient language design. The main artifacts of the language are rules, functions, metrics, time routines and characteristics. The concepts can be described as follows:

### Rules
Rules are logical and arithmetical expressions[15] that can be evaluated to Boolean values true or false. They are defined in the context of sensors and can be used to specify the desired behavior of a system. In our example the sensors are labeled with identifiers (i1, ..., i4).

A rule may contain logical operators like AND, OR, IMPLIES, NOT, IF-THEN-ELSE etc. and arithmetical operators like PLUS, MINUS, MULTIPLY, DIVIDE etc. An important concept to deal with complexity of such specifications is referencing sub rules or other language elements, like functions from a library (e.g., MAXIMUM, SUM, AVERAGE etc.). In our example the time routine StandardShiftOperation and the characteristics Characteristic1 and Characteristic2 are referenced. The referenced elements are specified self-contained in separate artifacts to enable reusability within other language artifacts, see Figure 5.

---

[13] OMG: OMG Unified Modeling Language Specification. Technical Report, Object Management Group (2003)
[14] Krahn, H.: MontiCore: Agile Entwicklung von domänenspezifischen Sprachen im Software-Engineering. Shaker, Aachen (2010)
[15] Rumpe, B.: Modellierung mit UML. Springer, Berlin (2004)

```
IF StandardShiftOperation AND i2 < 3
THEN Characteristic1
ELSE IF NOT StandardShiftOperation
AND i2 < i4
THEN Characteristic2
```
**Figure 5 An example of a rule specification**

After specifying a rule an automated analysis can be executed by the platform. For each rule (and sub element) a virtual sensor is created. A virtual sensor is analogous to real sensors an equidistant sequence of values. In the case of a rule a valid value can be true, false, missing or undefined. A result is missing if no context information (sensor data) is given. An undefined result means that no evaluation was possible. The resulting sequence has the same temporal resolution as the context sensors, for instance 15 minutes.

Functions
Functions are very similar to rules but they are not resulting in Boolean but numeric values. The context of a function is one or more real or virtual sensor. An example with two sensors is the function f(s1, s2) = s1 / s2 where s1 could be the indoor temperature and s2 the outdoor temperature. The function calculates the indoor/outdoor quotient, which can directly be used for visualization or can be referenced in other functions, rules or metrics.

Characteristics
Characteristics are a powerful concept to specify relations between two sensor dimensions. For instance the relation of indoor and outdoor temperature can be defined by the following characteristics.

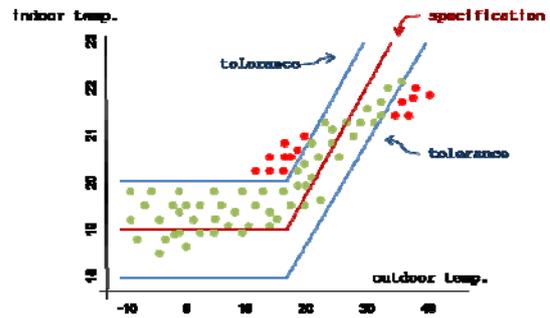

**Figure 6 An indoor/outdoor relation specified by a characteristic**

A point based upper and lower characteristic can be specified by a graphical editor, see Figure 6. The given example shows an indoor temperature on the vertical axis and an outdoor temperature on the horizontal axis. The points are concrete measured data from the two context sensors. The analysis matches each value with upper and lower bound. If an outlier is detected, the characteristic is interpreted as false for the related timestamp. Characteristics can be referenced in rules and functions.

Metrics
For a better understanding of difference between functions and metrics the following example shows the evaluation of concrete values. Functions, rules and characteristics are equidistantly process, what means that the value sequences are iterated from timestamp to timestamp and results are calculated by the given parameters. Figure 7 shows the evaluation of a function.

function: f(s1, s2) = s1 + s2 / 100.

| timestamp | 00:00 | 00:15 | 00:30 | 00:45 | 1:00 |
|---|---|---|---|---|---|
| s1 | 16.0 | 15.8 | 15.5 | 15.1 | 14.9 |
| s2 | 19.2 | 19.2 | 19.1 | 19.2 | 19.0 |
| result | 0.352 | 0.350 | 0.346 | 0.343 | 0.339 |

**Figure 7 An example of a function specification**

Compared to functions metrics are calculating values for a given time span. An example is the metric averageWaterConsumptionPerHour. Figure 8 shows the difference to the calculation before.

metric: averageWaterConsumptionPerHour

| time-stamp | 00:00 | 00:15 | 00:30 | 00:45 | 01:00 | 01:15 | 01:30 | 01:45 |
|---|---|---|---|---|---|---|---|---|
| s1 | 10 | 5 | 15 | 10 | 15 | 15 | 20 | 10 |
| result | 10 | | | | 15 | | | |

**Figure 8 An example of a metric specification**

A so called base metric consists of a context sensor, in our example this is a sensor that measures the water consumption in liters per 15 minutes. Additionally a base function can be added. Here we use the AVERAGE function. Last but not least a time filter can be added. We use PerHour in our example (we also support PerDay, PerWeek, PerMonth, PerQuarter, PerYear).

With this simple configuration mechanism a lot of standard metrics can be calculated. Additionally customized metrics can be used from a library which can be parameterized, e.g. the calculation of a standard deviation. Metrics are very helpful for management reports.

Time Routines
Time routines can be used to differentiate between several operations modes. An example for use is the specification of a public school building, where facilities should have different operation modes for weekdays, weekends or holidays. A time routine can be defined a set of time ranges. Values for year, month, day, hour, minute, second and weekday can summarized to a schedule. It is also possible to include or exclude additional time routines to specify exception, e.g., holidays. The following Figure 9 shows an example for a weekly shift operation.

| time routine: StandardShiftOperation | |
|---|---|
| Year | * |
| Month | * |
| Day | * |
| Day Of Week | Monday, Tuesday, Wednesday, Thursday, Friday |
| Hour | 8-17 |
| Minute | * |
| Second | * |

**Figure 9 An example of a time routine**

The * is a wildcard which means that for instance the year is not considered for evaluation. The Date "Thursday 1st of May, 2010 10:30:45" is matching the schedule and the time routine is interpreted as true. Contrariwise the date "Thursday 1st of May, 2010 07:00:00" is not matching and the routine is interpreted as false.

Ticket System
The ticket system is highly adoptable for each building. It can be used for our automatic and manual analysis of the building data. After our rule was evaluated by our system and failed because the facility is not running conform to the specification, we create a new ticket that holds all necessary information for the facility manager. So violated constrains or specifications from the design phase can be traced during the operation of a building.

Visualization
Our platform offer several types of visualization. We can use line plots, scatter plots and carpet plots. In Figure 10 you can see two sensors. The upper sensor is displayed as line plot, the lower as carpet plot. It is also possible to visualize the results of our metrics, rules and functions. These plots can be exported as images for further applications.

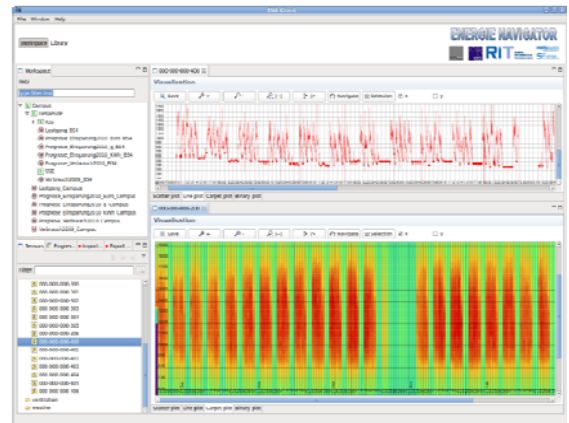

**Figure 10 Visualization of sensor information as line- and carpet plot**

Reporting
After the analysis we offer the automated generation of reports. These reports are customizable. It is possible to insert handwritten parts to the generated reports. So an energy expert can create a special purpose report which focuses only on the heating facilities of a building that consists only of the important sensors, facilities, rules and visualized plots. This report can then be reused for a specific building. The expert can make additional comments, e.g. to give suggestions what should be changed for the operation of the heating facilities. These comments are saved independent of the generated report. If the report is regenerated, e.g. after one month new operational data of the building is available; the original handwritten comments are still there.

4. APPLICATION
The main target of the application is to provide a tool to describe building functions and to control them with the same document. The standard procedure follows the following steps:
1. In the design phase the building and its systems will be defined usually by the

design engineer. All relevant rules for design AND operation are designed using a system scheme and the active functional description. Additional information can be entered in conventional text editors.
2. In the construction phase the necessary data points from the building automation system will be added usually by the manufacturer of the BAS.
3. In operation the BAS will be linked with the Energy Navigator for example via OPC.

Figure 121 shows a screen shot of the expert tool describing the consecutive steps in a general application.

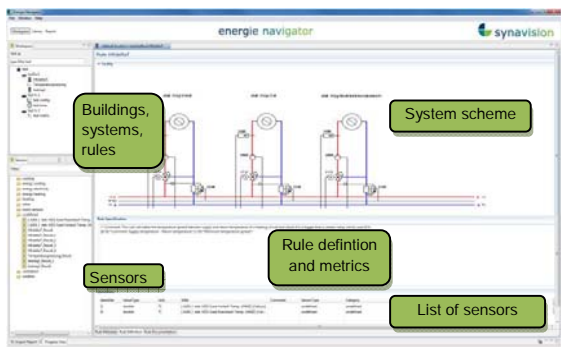

**Figure 11 Expert tool and application process in the design phase**

A reporting engine or ticket system will analyze building operations and automatically create reports or create tickets for a continuous workflow. Reports can use common visualization types and can be amended with text, see Figure 12.

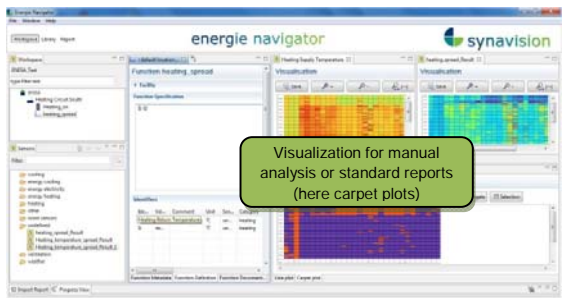

**Figure 12 Expert tool and examples of visualization as basis of reports**

Since the tool is web based it can be used in multiple ways. In the European project *Best Energy*, funded by the European Commission, the backend is used as data source for information screens and a project website providing information on energy consumption to users of a building at the Technical University of Braunschweig/ Germany, see Figure 13.

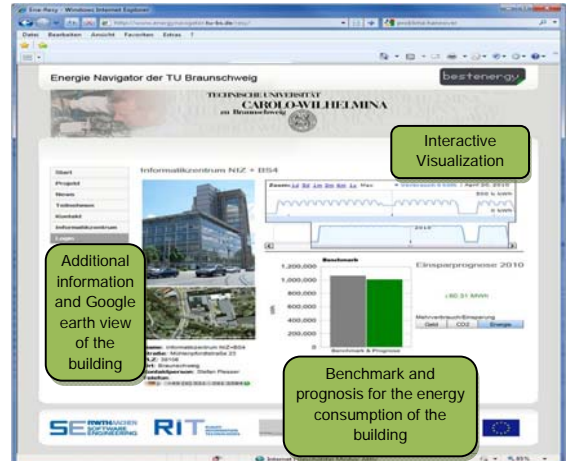

**Figure 13 Public Website at Technical University of Braunschweig (Project Best Energy)**

The website is intended to increase public awareness of energy consumption in combination with a system of incentives for the users letting them participate in energy cost savings.

5. CONCLUSION

The Energy Navigator is an innovative approach to improve the quality of building functions. Being implemented already in the design phase of buildings the process of functional quality management can make use of the design engineers know how. Furthermore it provides means to maintain an up to date documentation even after a commissioning process with multiple operational adjustments.

For potential products and market success the concept of the Energy Navigator has big advantage: a functional description for building systems has to be created anyway during the design process. By using this description as a reference model in monitoring the new approach does not require an extra model. The integration into the life cycle of design and operation and the economic potential could help to make necessary monitoring a more common service in the near future.

ACKNOWLEDGEMENT
The project is funded by the German Ministry of Economics and Technology (FKZ: 0327444A).
The public monitor website is used in the project "Best Energy" ICT Policy Support Program (ICT PSP) Grant Agreement No 238889 funded by the European Commission.